PREPRINT

# Are Smart Contracts and Blockchains Suitable for Decentralized Railway Control?

Michael Kuperberg[†], Daniel Kindler[‡], Sabina Jeschke[§]

**Abstract.** Conventional railway operations employ specialized software and hardware to ensure safe and secure train operations. Track occupation and signaling are governed by central control offices, while trains (and their drivers) receive instructions. To make this setup more dynamic, the train operations can be decentralized by enabling the trains to find routes and make decisions which are safeguarded and protocolled in an auditable manner. In this paper, we present the findings of a first-of-its-kind blockchain-based prototype implementation for railway control, based on decentralization but also ensuring that the overall system state remains conflict-free and safe. We also show how a blockchain-based approach simplifies usage billing and enables a train-to-train/machine-to-machine economy. Finally, first ideas addressing the use of blockchain as a life-cycle approach for condition based monitoring and predictive maintenance in train operations are outlined.

KEY WORDS

## 1.  Introduction

Unlike car traffic, most mainline railway operations have technical frameworks to constantly enforce strict safety procedures. These frameworks are designed and implemented to withstand an operator's failure or even death. While the scope of the frameworks differs between countries, the state-of-the-art implementations such as ETCS[1] include the constant upkeep of "safe blocks" (to prevent collisions even in case when one of the trains comes to an unexpected stop or derails), emergency braking if a red signal is passed (or if the human operator does not react within a specified time), detection of train decomposition, and variable speed control. Trainside and lineside IT components work together to achieve these goals.

Beyond the safety-guaranteeing frameworks, railway operation requires live dispatching: in addition to schedule-based passenger/freight trains, dispatching must accommodate ad-hoc traffic, deviations, construction-caused alterations, equipment failures etc. Despite advances in conventional and AI-supported decision making, a lot of this work is still performed by humans, i.e. experienced dispatchers. Dispatching and safety

---

[†] Michael Kuperberg (michael.kuperberg@deutschebahn.com) is the Chief Blockchain Architect of DB Systel GmbH, the IT provider of Deutsche Bahn AG

[‡] Daniel Kindler (daniel.kindler@deutschebahn.com) is the Managing Partner Blockchain & DLT Solutions – DB Group Business Segment Infrastructure of DB Systel GmbH

[§]Sabina Jeschke (sabina.s.jeschke@deutschebahn.com) is a Member of the Management Board Digitalization & Technology (T) of Deutsche Bahn AG

frameworks are usually partially decentralized for large railway networks: they are split into regions so that size and complexity are manageable – very similar to air traffic control operations. Over time, such operations have been electrified, electronical equipment has been introduced, and the newest equipment generation relies on digital[2], semi-detached infrastructure elements (switches, signals, occupancy sensors) as well as on in-cab signaling.

Dispatching and safety frameworks are complex, heterogenous and very costly (with an invest of between 100,000 and 300,000 € per km[3]), developed over many years and with lifecycles of several decades. Despite attempts at standardization, both interoperability and vendor lock-in pose an ongoing challenge. Additionally, such frameworks have historically been nation-specific; international standards such as ETCS require substantial investment in hardware and software during a transition phase. Ultimately, this should help overcome the heterogenous patchwork of country-specific standards in place throughout the EU: cross-border train operation often requires additional training, multi-system vehicles (at a higher cost), or changing trains/staff at the border.

Still, railway traffic management (incl. dispatching and safety/security) is not part of ETCS; such functionality is being designed as part of ERTMS[4], which is the overall initiative that encompasses ETCS and GSM-R. Infrastructure utilization is the key to lower operating costs, and customer satisfaction is strongly correlated with punctuality and density of service. Thus, some progress has been made on improving track utilization (e.g. using "moving blocks"[5]), in addition to improvements in safety. Despite these advances, the principles of railway operations remain largely the same: control center is "the master" and the train is "the slave". This resembles the mainframe design patterns, where the trains are only the "terminals".

This hub-and-spoke pattern remains in place even for cutting-edge "autonomous train operation"[6]. Thus, the mainstream approach is a "slow evolution", mandated by the backward compliance in large networks but also by the intrinsic interests of the manufacturers and investors. However, there are situations where seamless train-to-infrastructure and train-to-train contracting would lead to improvements: trains could dynamically negotiate and "sell" a timeslot, automating redispatching in a rational, market-driven way. Potentially, passengers can request unscheduled stops (and bid/pool for them), and unpredicted construction or extension of maintenance shutdown periods could be propagated across the network. Additionally, trains could self-report operation-impacting defects (such as overheated axle bearings or derailments).

Another potential for improvements exists in the "back office" area: many national rail networks provide "open access" to competing passenger and freight railways, which pay regulated fees for infrastructure usage (tracks, stations, energy supply). Likewise, the "back office" sells ahead-of-time access rights since network access is strictly controlled to enable timetabled train operations to maintain their quality of service. Using a blockchain, *both* the sale/allocation of access resp. usage rights *and* the actual payment for them would happen transactionally and instantly, in one system rather than in several. In the area of maintenance and servicing, currently, maintenance windows, maintenance intervals, and maintenance plans are designed to ensure the safety of the system. The use of blockchain coupled with the ever-evolving sensor technology of trains or infrastructure components

**2**

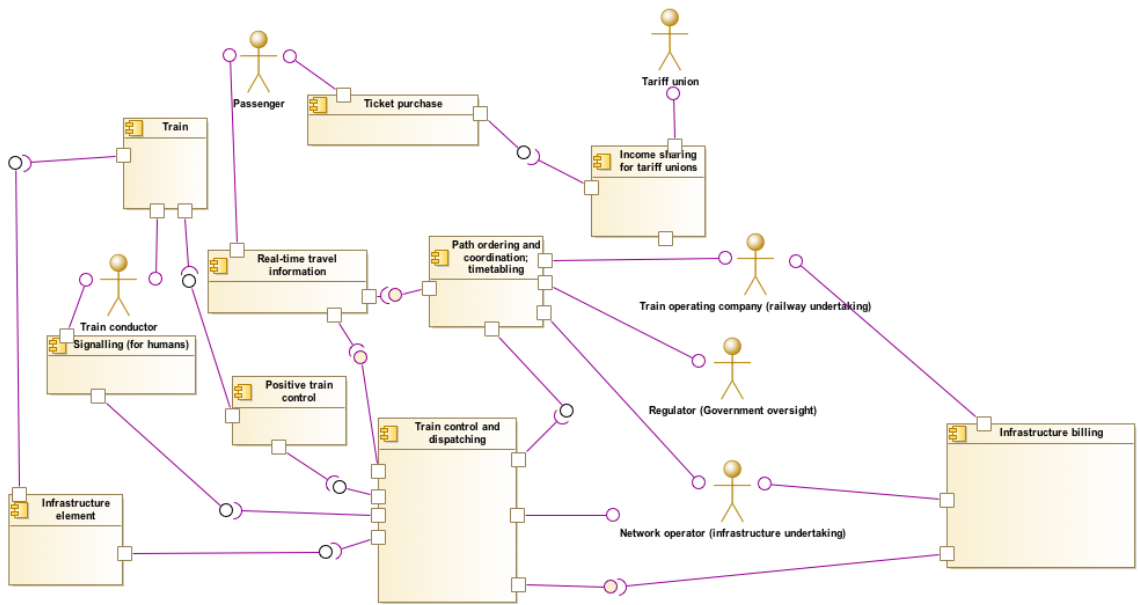

**Figure 1: some IT building blocks in deregulated passenger train operations**

can be seen as the basis of modern maintenance, in which the individual components independently register their requirements. The rollout of such on-demand maintenance then includes e.g. also the automatic ordering of spare parts or the provision of special teams.

In Figure 1, we show the most essential parts of this "ecosystem". Each shown part exists in several instances: there are sovereign rail networks both at national level and regional levels. In reality, there are additional layers (e.g. procurement, malus/bonus processing, HR, maintenance planning and management, construction, governmental oversight, insurance) which we do not detail here. Ultimately, we are working towards a modular, API-driven multi-modal "Transportation Operating System" where all interactions (from ticket purchase to subcontractor payments) are handled on the basis of a consistent, trusted, replicated ledger "data base". Some of the functionality shown there has already been showcased by us in previous works: e.g. blockchain-based station usage billing [7] and blockchain-based revenue sharing [8]. Throughout these modules, privacy, reliability and performance (throughput, latency) remain an ongoing challenge.

In this paper, we investigate a disruptive approach to train control based on Distributed Ledger Technologies (DLTs) that thoroughly rethinks the involved roles: a DLT becomes the trusted "single truth" for both the current state of the networks and for the "future infrastructure reservations" which are the results of pre-scheduling and ad-hoc planning. The ledger data is replicated across ledger nodes, providing fault tolerance and reliability. Furthermore, smart contracts serve as "gatekeepers" to the state changes (e.g. new reservations or cancellations) and the participating nodes cross-verify these changes, implementing a consensus mechanism that ensures consistency with specified rules - across nodes. We show how a prototype implementation works, and where additional research is needed. As part of the paper, we discuss the challenges to this approach, and compare it to existing implementations. We also show how it can be developed to a modular "Transportation Operating System".

The remainder of this paper is structured as follows: Section 2 shows the full complexity of the problem and how we have defined the scope for the initial proof-of-concept (which was implemented using a training facility used by Deutsche Bahn AG). Section 3 presents the architecture of the solution and explains the choice of the used technologies. Section 4 describes some important implementation aspects (incl. the use of the Ethereum blockchain stack) and the 'lessons learned' that we gathered during the PoC. Section 5 analyzes related work and how it compares to our results. Section 6 concludes and provides an outlook, together with the planned next steps.

## 2.   Work objective, scope definition and assumptions

Our research objective is to enable trains (resp. train operators) and infrastructure elements to be first-level, *active* and self-aware participants in railway control systems. Active participation by vehicles builds on access to a trustworthy, up-to-date view of the network use - past, current and planned. Such active participation includes wayfinding (both ad-hoc and in-advance), booking and actual usage of the infrastructure, as well as interaction with lineside equipment (e.g. setting a switch into the correct position) and with other vehicles. As part of our objective, we want to establish an authoritative data repository that is audit-proof/tamper-proof (through write-once-read-many semantics, aka WORM) and which relies on open-source, security-assessed COTS software components rather than on proprietary technology.

The write access to the data repository is to be protected by "gatekeepers", which ensure that only secure entries can be inserted (e.g. no two trains are located in the same block at the same time). The WORM semantics mean that data cannot be overwritten (not even by consensus). Some nodes may choose to store only a certain, regulation-imposed backlog of past data to keep the data amount manageable.

Train localization and integrity checks (detection of decomposition) are assumed to be available; there exist established technologies for this (such as trackside equipment, GNSS, Differential GPS[9] resp. axle counters[10]). Trainside safety/security measures (such as a forced emergency stop when trying to enter an occupied block) would access the state information (e.g. block occupancy) stored in the proposed, new system - currently, such information is transmitted by trackside equipment/IT (e.g. LZB, magnets in PZB/Indusi) resp. by GSM-R (as in ETCS Level 3). Therefore, the physical reality and the IT representation are matched: the infrastructure elements and the trains have "IoT digital twins".

It is imperative that a digital twin and its physical counterpart are "mutually reliable" for both state representation and state changes. For example, the physical switch must be reliably "locked" except during state changes, and may not change its position without having been instructed by the digital twin to do so. If a switch position is modified by brute force (e.g. through sabotage), the malfunction must be detected (e.g. by an appropriate circuitry) and the disruption must be represented in the IoT digital twin, preventing an accident. We assume in the following that infrastructure elements and the synchronization between the IT representation and the physical entity is available and reliable, i.e. solved outside of our work.

Likewise, we assume that the trainside safety aspects of the (automated or human-operated) train operation are maintained. However, they information they rely upon are not provided by the centralized control center, but rather by the blockchain-based network. For example, trainside emergency stop is auto-activated if a train enters an occupied block.

It is a part of our approach that a train pays the infrastructure usage directly to the infrastructure element (e.g. to a track stretch). This means that the infrastructure element has a way to receive payments, and its owner can administrate those payments. Payments can also be bonuses (or fines), or "back office" tasks as in Figure 1. To enable such a M2M economy, we propose to use blockchain wallets (which are effectively PKI keypairs, cf. Ethereum), where the balance of a wallet is stored on the blockchain ledger. In this paper, we do not discuss how the payment is integrated into the planning/management/control processes.

## 3.  Solution Architecture and Employed Technologies

Conventional train control systems are usually centralized in two ways: logically and technically. Logically, there is just one business entity (the "infrastructure operator") running the control system, and it is the only party that has full access rights (incl. write permissions); it may or may not allow read-only access for train operators at its own discretion. Technically, the control system is usually centralized because cost factors do not encourage a multi-node/multi-location setup. Additionally, a "hot standby" or even "active-active" setup means that data must be replicated successfully and completely *before* it can be considered as "written through"; this may increase latency and strain the data links between the locations.

Increasingly, infrastructure malfunctions lead to compensation claims from the train operators, which themselves have to pay compensations for delays to customers. This forms a monetary driver for further fault tolerance in railways operations; additionally, the ongoing progress in hardware performance-to-cost ratio encourages the "design for failover" approach even in the light of the additional implementation costs compared to the simple setup.

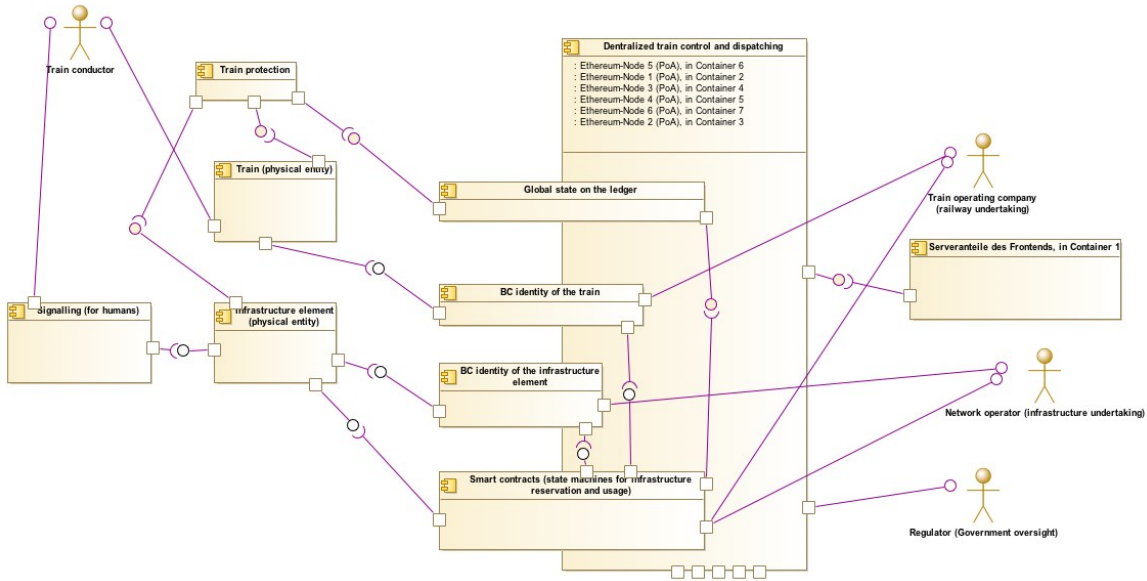

**Figure 2: architecture of the prototype**

Therefore, our architecture (as shown in Figure 2) is primarily based on *technical decentralization*, i.e. on multiple nodes. The blockchain identity of the train is the actor which orders paths (i.e. performs reservations of an infrastructure element for a given timeframe); it interacts with the blockchain identity of the infrastructure element via the smart contract which is the gatekeeper (ensuring consistency and a safe global state). The smart contract is the entity which changes the "should be" part of global state. When it comes to the "is" global state, the physical world is the "leading truth"; the state on the blockchain is mirroring the "physical truth". The train protection component does rely on the "is-state" but consults the future state as well.

When it comes to logical decentralization, the situation is more intricate. Logical decentralization means that the nodes belong to multiple parties. This immediately poses the questions of authority, agreements, responsibility and liability. When it comes to security and human life, strong authorities are the traditional choice. Implicitly, a single authority means a single (central) responsibility. In aviation, the pilot and the co-pilot are an example where strict rules of authority in a multi-party setup are used to prevent a stalemate (standoff), as there is no arbiter to act as an intermediary between the parties.

Logical decentralization is inherently more complex than logical centralization, as it needs to address the situation with failing/unreachable nodes, the meaning of dissenting minorities, party splitting and unstable behavior (cf. the "Byzantine generals" problems and the associated body of research). At the same time, large-scale networks with decentralized decision-making have appeared and maintained operation, e.g. the public Bitcoin and the public Ethereum blockchains. Such networks succeed in horizontal scaling, a working set of rules for consensus (which is a systematic decision-making using defined majority rules), and in fault tolerance. At the same time, achieving suitable QoS and performance (latency, throughput, predictability) while maintaining scalability remains challenging in decentralized DLTs and blockchains.

Ultimately, our solution architecture is technically suitable for *both* logically centralized *and* logically decentralized setups: since we use a *private-consortial* (non-public) blockchain as the "engine" to run our algorithms and store our data, additional nodes can be added and additional parties can be onboarded. Note that for our architecture, there is no need for separate "oracles" that supply externally-sourced information. For timer-triggered events (e.g. checking whether a reserved element can be released since the reservation has expires and the train has vacated the element), reliable solutions are rather straightforward to implement.

It is important to stress that reservations (and payments) are handled in a peer-to-peer fashion between trains and infrastructure elements (such as switches); the IoT twin of the infrastructure element is in control of the element's blockchain wallet. Blockchain, on the other side, is the event bus and the recording ledger, but it is *not* a first-level, self-aware, individually-acting entity with own interest. At the same time, the participants of the blockchain network safeguard the outcome of the peer-to-peer transactions, because these transactions (e.g. admitting a train into a track section) are security-relevant and affect all peers – not just two.

State changes on a blockchain consists of several steps. Independently of the used technology, the three core steps on the "happy path" are transaction proposal, transaction validation and the replication of the validated transaction; there might also be technology-specific steps such as ordering of the transactions. The transaction validation is the most intricate step: it is where "mining" could be included to combat spamming resp. to introduce incentives to "compute" the block. In our situation (private network), transaction validation is *the* trust-intensive step. The important design question to answer here is: how many network participants have to vote in favor of a transaction to validate it? The possible answers re "at least n" (n≥1), "at least 51% of all participants" etc. Our architecture is very flexible w.r.t. the validation algorithm, and we have employed Proof-of-Work (with minimal complexity), Proof-of-Authority and also Proof-of-Stake, while adjusting the consensus thresholds.

For the implementation of the state-keeping and of the smart contracts, we used the open-source Ethereum blockchain (geth) in a private/consortial setup, as described in the next section.

## 4. The Prototype Implementation of the Blockchain-based Control Core

To validate our approach, we looked for a physical system that would be as close as possible to a real-life mainline railway. At the PoC stage, using a real railway would incur risks that could affect human lives, and a "secluded" full-stale test setup was not available. As a replacement, the Darmstadt training facility for infrastructure operators ("Eisenbahnbetriebsfeld Darmstadt" near Frankfurt am Main, known as EBD, cf. Figure 3) was a very good opportunity:

- The facility is a long-standing joint venture between a research university (Technische Universität Darmstadt) and the Deutsche Bahn AG (resp. its DB Netz AG subsidiary)
- EBD is actively being used for academic teaching and railway research, having been employed in a significant number of projects[11]

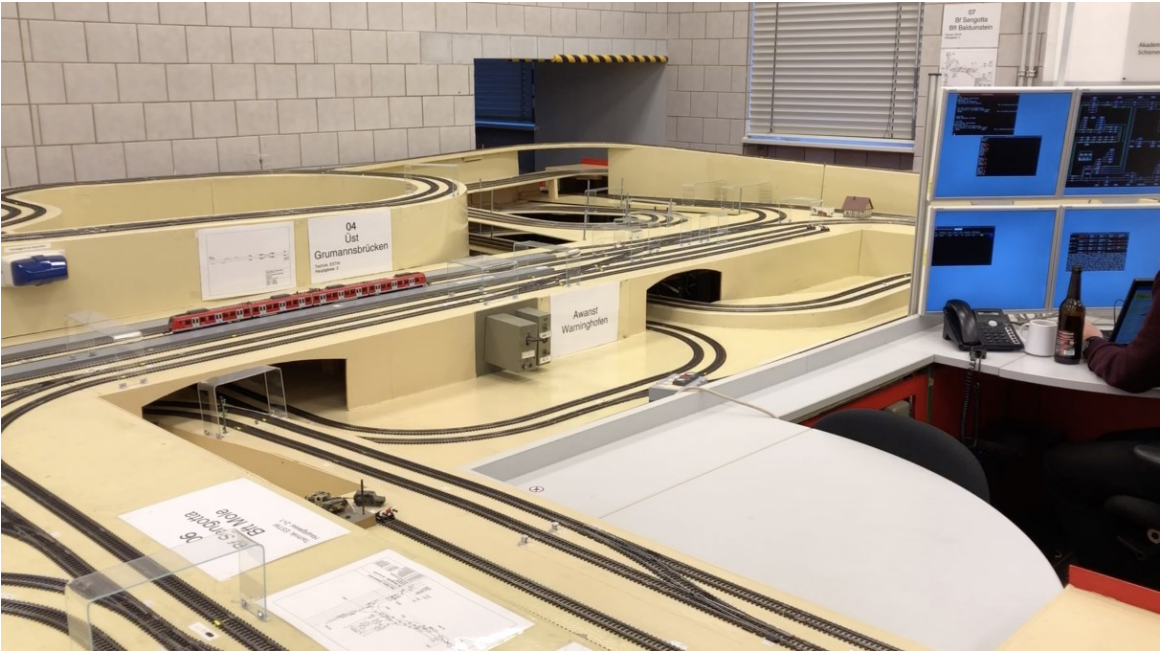

Figure 3: EBD validation setup (excerpt of the railway layout)

- The EBD includes different generations of railway control equipment (from mechanical to electronical), so it is clear which functionality of it our approach would replace
- The facility includes a large, complex model railway layout that is fully digital and includes a variety of train material, working signals and switches, as well as "section occupation" information as well as "vehicle location" functionality (both can be used through documented APIs and over established network protocols)
- The existing security/safety/localization software and hardware in the EBD would remain as-is for our undertaking, while the train control and dispatching were to be engineered using blockchain technology, following the architecture described above.

We have decided in favor of the ready-to-use EBD and the against a custom-made scale model railway even though the model railway would be transportable and thus better suitable for on-premise demonstrations. The reason to start with the EBD is that the reproducibility and trustworthiness of our results from the perspective of domain specialists (both non-IT and IT) is strengthened when using a validated, established third-party setup as a foundation. Our implementation supplies train control and dispatching functionality that works "on top of" the EBD-provided interfaces. Our contribution includes additional detailed checks not performed by the EBD software itself: e.g. we ensure that switches are *always* in the correct position.

As the EBD is a scale model of the current, centralized mode of railway operations, the EBD's model locomotives cannot be active participants of the blockchain-based approach: they do not have computing power or communication facilities, even for lightweight API-based access to blockchain nodes. In fact, even the trainside safety aspects (such as emergency full-stop if a red signal is passed) are virtualized inside the EBD by the central control logic of the model railway. To enable humans (i.e. train drivers) to "reserve" paths (i.e. necessary sequences of infrastructure elements for the specified time), we provide a graphical UI which visualizes every step, from wayfinding over payment to the actual usage. For automated operation, the underlying input (departure and arrival - points and times) may be provided by passengers or by operators from the train-owning railway undertaking.

In this context, by now one known disadvantage of using Ethereum is that "out of the box", it is an *unpermissioned* (permissionless) blockchain: *theoretically*, everyone (within the consortium) is free to set up one or several nodes and to participate and *potentially*, every participant (and every node) is equal-righted. Technically, there is no PKI-based authentication and authorization in Ethereum (e.g. unlike in Hyperledger Fabric), leading to anonymity/pseudonymity.

However, there are two levels of protection in place for our implementation: the network-level protection restricts access to nodes based on a whitelist. Additionally, at the level of smart contracts and assets, Ethereum's concepts and the Solidity language for smart contracts provide built-in mechanisms of ownership, delegation, and custom-defined permissions for custom-defined assets and contracts.

For the initial implementation, we have chosen Ethereum over e.g. Hyperledger Fabric because of the built-in wallet functionality and because there are ready-to-use implementations of Proof-of-Work (PoW), Proof-of-Stake (PoS) and Proof-of-Authority (PoA) consensus algorithms. At the same time, we fully understand that the privacy-preserving concept of "channels" (e.g. in Hyperledger Fabric) is more suitable, and blockchain-native tokens and assets can be added to most ledger implementation. Other products may also provide more choices w.r.t. programming languages for smart contracts, while Solidity is the only smart contract programming language available in Ethereum. As other ledger technologies (e.g. Hashgraph, R3 Corda, Hyperledger Burrow, Neo, …) mature, we will re-evaluate this choice.

The UML sequence diagram in Figure 4 shows the steps involved in a successful reservation of a multi-resource travel path. The release of the infrastructure elements can be "as quick as possible" (as soon as the train has passed them, which improves efficiency), or can be triggered later (incl. "implicit release" when the reservation expires, provided that there is no active usage of the infrastructure element).

The route-finding on throughput-constrained, weighted/priced and time-aware graphs is already solved by several libraries and products. In our approach, the current and future reservation are stored on a distributed ledger; at the time of writing, there were no libraries that would expose such route-finding functionality *directly* on ledger-stored graphs. Thus, we have used a rather simple JavaScript library for finding potential routes, but JGraphT[12] is one of the more elaborate candidates. *Available* routes (a subset of *potentially possible* routes) are then determined on the basis of the ledger-stored reservations.

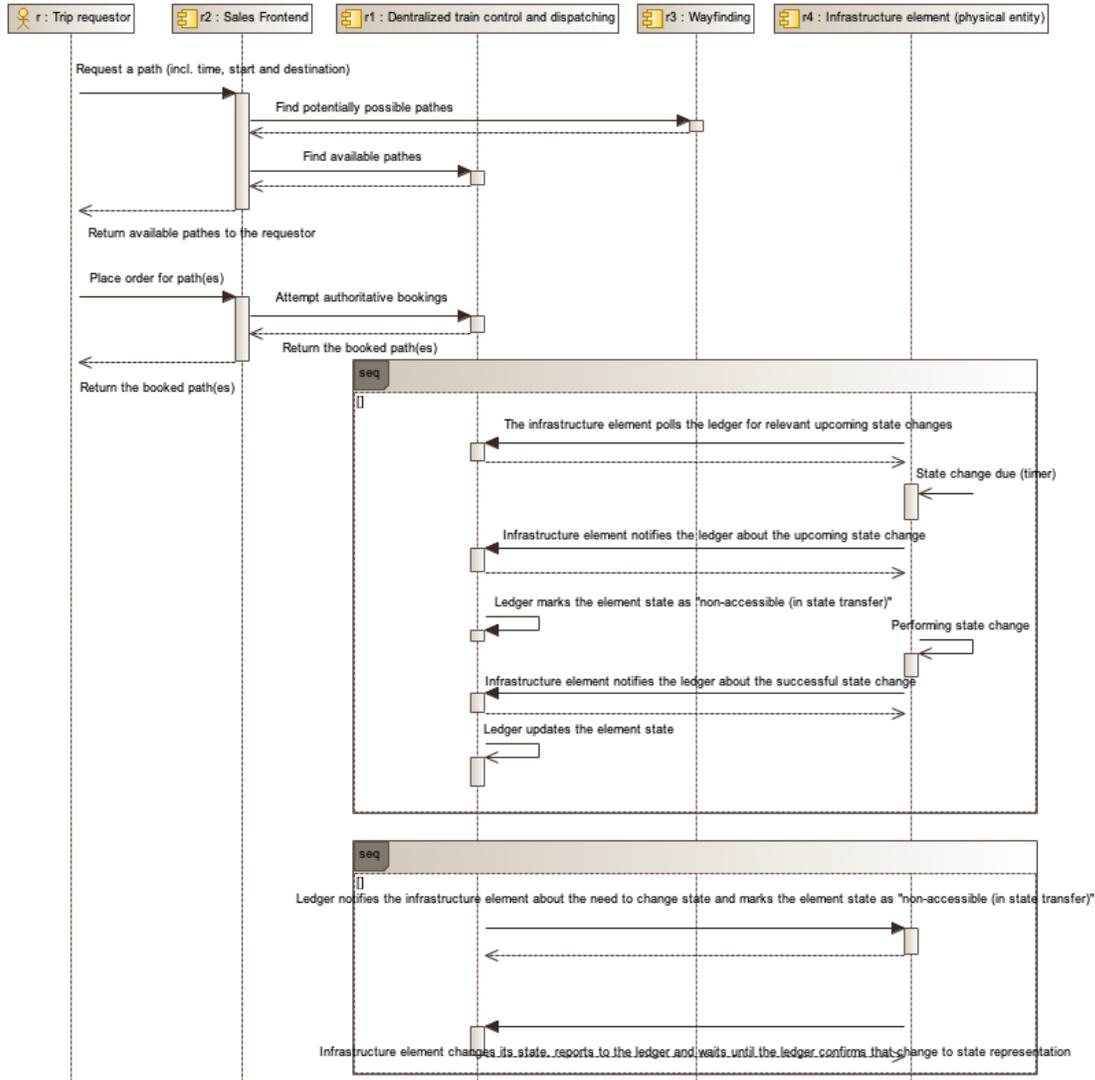

**Figure 4: sequence diagram of the path reservation logic (excerpt, simplified view)**

Both the graph search and the subsequent reservation are subject to transactional concern such as atomicity, consistency, isolation and optimistic/pessimistic locking. For example, if the set of the potential routes (which are found by a train) is not locked, there is a probability that another train will book a part of that potential route. Then, by the time that the initial train decides to book a potential route, it may have become unavailable.

Such problems are not DLT-specific and solved in different ways (optimistic, speculative, pessimistic, time-constrained locking) in existing systems. To keep our prototype implementation simple, we did not implement pessimistic transaction locking but instead act in an optimistic way: we accept the possibility of a "booking failure" if the selected potential route has become unavailable in between (some or all segments, that is). In such a case, the wayfinding and reservation attempts are simply repeated.

Likewise, for the initial implementation iteration, we chose to live with possible side effects of non-atomic behavior, when a route with multiple resources is booked: if the first resource booking(s) succeed but a following resource booking fails, the already-booked segments of the chosen route are "released" (un-reserved). Then, the train has to re-try again.

Including locking and transactionality into the booking contract is of course possible, but we decided to do it once the technology stack for the next version of the approach has been re-evaluated: we are investigating ledgers such as Hyperledger Fabric, R3 Corda and others because they offer sub-groups ("channels") so that not all information is broadcast to every node in the network.

Railway-operating software is subject to strict norms (e.g. EN 50128, 50129) and SIL (safety integrity level) procedures. In addition to rigorous testing, formal verification may be required to prove the suitability of both platform components and application components, as a prerequisite to certification. These tasks will be part of our future work.

Some Blockchain implementations do not prohibit or (as Ethereum) even explicitly support *forks*: branches which make the ledger a non-linear, tree-like structure. A fork marks a "split" where state changes are non-serialized and two branches *can* contain two conflicting statements about the same item. Forks can be intentional [13]; but fork-like situations can also happen if the network splits in two parts with no interconnections between them: each network part develops its own version of the ledger. Obviously, forks (resp. a nonlinear blockchain *without* consistency checks) bring the risk of ambiguity – something that must be avoided in train control.

When we chose Ethereum for the prototype implementation, we were aware of the possibility of forks on the public ledger – and even as we are using a consortial, non-public setup, network partitioning cannot be fully excluded. As part of our future work, we plan to conduct a deeper evaluation of other DLT technologies and products, and avoidance of forks as well as the detection/avoidance of network partitions will be a key evaluation criterion.

While the Ethereum blockchain has the "replicate everything on each full node" principle, it is possible to reduce the data load (e.g. through *sharding*) and participants can use asymmetric cryptography to encrypt private information passed over an unencrypted medium (DLT), given that the public keys (for the encryption part) are already available as part of the wallet. Other enterprise-grade DLT/Blockchain stacks (such as Hyperledger Fabric from the Linux Foundation) offer further facilities for privacy scoping, e.g. channels.

## 5. Related work

Decentralization in railway control has been addressed in multi-agent research ([14], [15] and others), incl. comparisons of performance between centralized and decentralized scenarios. However, these approaches have neither used a tamper-proof, transparent ruleset (as the Ethereum smart contracts that we used), nor did they use a tamper-proof "full history" approach (as we do with the Ethereum-based distributed ledger/blockchain). Beyond the concepts, none of these approaches has been validated in a real-world training facility.

Outside the railway industry, autonomous vehicles are currently not able to pay for their infrastructure usage (such as tolled highways, bridges, or parking facilities) "on their own" (in an unsupervised manner). While there are proofs-of-concepts[16] and interest groups[17] with the focus of car-to-infrastructure payments, they are neither targeting ahead-of-time reservations nor do they cover limited-access, restricted-capacity infrastructure that is the cornerstone of railway operations.

Our approach includes the enablement of trains to establish trade relationships with each other, e.g. to enable unsupervised monetary compensation from a delayed express train to a freight train when the latter yields its priority so that the express train can reduce its delay. The communication part of this vision can be compared to "car-to-car communication" [18], which relies on technologies such as 5G. In contrast to such work, our contribution focuses on the protocols and on the contents of such M2M communication; strictly speaking, the blockchain and its consensus introduces an intermediary layer.

Using blockchains (or distributed ledgers) for machine-to-machine payments has been studied and demonstrated on several occasions. This is encouraged by the native (crypto-) currency capabilities and token concepts of the underlying technologies, such as ERC20 in the Ethereum ecosystem or NEP-5[19] in NEO. Still, none of these approaches proposes or even implements the solution to the systemic constraints of mainline railway operations.

Non-DLT solutions for WORM data storage include dedicated hardware [20] and software/cloud[21] designs, modifications of existing file systems[22], or data archival[23]. Some database products support replication and "hot standby" / "cold standby" modes[24], usually with a cluster-based distribution. In a similar way, Master Data Management systems[25] are concerned with data replication and synchronization. These solution types incur a vendor lock-in (due to proprietary software) and are subject to the "deleted by the master administrator" problem. Additionally, they do not scale across enterprises, across thousands of nodes, and do not provide the auditability of the "rule source code" and "rule execution" as is the case with the blockchains/DLTs.

Our vision includes "sovereign" infrastructure elements such as switches which are adhering to the ledger rules and events (from smart contracts). There are different approaches to build railway IT infrastructure in a more networked/peer-to-peer way[26]. However, none of them covers M2M payments and train-to-train economy.

More modern paradigms to switch controls and automated dispatching are to be found in "turnkey" systems for urban/suburban networks[27], often with CBTC[28] (communication-based train control). These networks have central management facilities which control both trains and infrastructure in an integrated way, incl. peak traffic management and ad-hoc addition of trains. However, so far, none of such systems has been applied as "in-place upgrade" to a mainline network, as this would require large-scale adaptation of trains and infrastructure and also the (costly) coexistence of the "old" and "new" systems, as larger networks cannot be upgraded in one "big bang" step.

A certain overlap with our approach can be found in newer signaling and control systems such as ETCS[1], where the train is more aware of its surroundings. However, even with ETCS, the physical train is obtaining certain constraints (maximum speed, …) in a *passive* way and for safety reasons, the human train driver can only operate within those constraints.

From the economist's perspective, our proposal can be classified as a "decentralized transparent free market economy with strict rule enforcement", whereas the existing mainline approach is more "centralized market economy with central resource allocation in a non-discriminatory manner" (in Germany, the deregulated railway ecosystem includes a regulatory/oversight authority which is tasked with ensuring the open-access policy set forth in the law). There is a substantial body of work investigating the (dis-)advantages of

**12**

central coordination in comparison with decentral, "peer-to-peer" trading. However, such research is rarely applicable to the allocation of scarce resources (as it is the case in mainline railways) in combination with safety constraints that are inherently complex to include in a market model/simulation.

## 6. Conclusion and Future Work

In this paper, we have introduced blockchain-based decentralized railway control and its prototype implementation, which combines traditional safety principles with cutting-edge IT technology. Trains can determine possible routes and book them directly (both ad-hoc and in advance), based on transparent and binding smart contracts which ensure conflict-free resource booking. At the same time, our approach coexists with centralized batch planning that produces long-term timetable and which integrates pre-planned construction sites - across railway undertakings.

As the ledger lists all binding recordings of the current status and the advance reservations, a train-to-train economy without the risk of double-spending, deniability or repudiation can be established. The trains, infrastructure elements or other vehicles can participate in the asset exchange in an autonomous way, or using "human proxies" from the train-operating companies resp. staff; the machines become proactive members of the transportation network. Essential parts of the specialized hardware, software and infrastructure – interlocking, signals and control centers – can be streamlined, merged and redesigned to be more fault-tolerant.

In combination with the ledger immutability and the unambiguous assignment of actions and assets to an identity, a virtual but trusted identity and a trusted curriculum vitae (CV) can be created and traced. Every part of the ecosystem (whether static or moving) becomes identifiable and possesses a history. Such data opens up new opportunities, e.g. in digital and predictive maintenance and in cross-enterprise asset exchange. Also, certain necessary administrative processes can be greatly simplified by using the virtual wallet for immediate monetary transactions. Vehicle usage and infrastructure services can be billed automatically; charges and reservation fees are paid seamlessly step by step, in the same system. Switches and other track elements as participants of the blockchain can receive payments.

The prototype implementation has proven that the concept is viable, as demonstrated by the in the EBD training facility control center in Darmstadt. As the next steps, we plan to employ the blockchain in combination with multi-agent systems (MAS) and IoT middleware to study the effectiveness of our blockchain-based, autonomic peer-to-peer economy for the railway operations. Aspects such as the trading of priorities, additional stops for ensuring trip connections and passenger transfer, surge pricing, international journeys or seamless on-demand offers in passenger and freight transport are on our agenda.

Another vector of our interests is a trial with full-scale real-life equipment, e.g. in marshalling yards with remote-controllable unmanned mini-shunters or on secondary lines with manageable complexity and little traffic. The focus would be to bring the entire approach into an operational state and to investigate scalability and performance. For this scenario, we aim at a close interaction and collaboration with the government-devised regulation authorities.

**13**

In parallel to the use of blockchain for decentralized control models, the power of the method in railway operation lies in the field of maintenance and service. These tasks are highly labor-intensive, cost-intensive and time-consuming. A huge number of existing delays and other shortcomings are directly or indirectly attributable to this area. A comprehensive digitization of this field includes - in addition to the sensory recording of the states of all components in real time - the proof of the execution of repairs and maintenance as well as the documentation and traceability of the single steps. The digital twin of a train or the digital twin of a component of the infrastructure is therefore time-dependent. Its dynamics can be understood as an autogenerated CV of the corresponding component realized by blockchain.

In addition to Ethereum, other frameworks and consensus mechanisms will be evaluated and, if necessary, integrated based on specific requirements (such as robustness against network partitioning). Also, we plan to study whether using state channels (to reduce the workload on the chain) would be suitable without compromising safety, security and reliability.

## Acknowledgements

Michael Leining, Michael Ziemek and Christian Klug provided valuable specialist feedback and helped set up the project. Karl Schöpf organized the funding for the project and contributed to the project scoping. Holger Kötting and Jan Niklas Dornbach were very helpful in making use of the software and hardware of the *Eisenbahnbetriebsfeld Darmstadt* and also provided us with documentation and code samples. Niklas Armbruster, Simon Dosch, Christoph-Samuel Pitter and Frank Polster contributed to the implementation. Steffen Ortolf contributed to resource planning and project management. Patrick Charrier challenged the design and provided valuable assistance at different stages.

## Author Contributions

MK designed the architecture, led the implementation, researched related work and prepared the manuscript (80%).
DK supplied additional domain knowledge and research material (10%).
SJ contributed the aspects of maintenance and trusted CVs for the trains and their parts (10%).

All WWW references were last accessed and checked on October 25th, 2018.